\documentclass[prb,fleqn,superscriptaddress]{revtex4}

\usepackage{graphicx}
\usepackage{subfigure}
\usepackage{dcolumn}
\usepackage{bm}
\usepackage{longtable}
\usepackage{amssymb,amsmath}
\usepackage{multirow}
\usepackage{tabu}
\usepackage{chemarrow}
\usepackage{amssymb,amsfonts,amsmath}
\usepackage{color}
\usepackage{float}
\usepackage{booktabs}

\begin{document}
	\bibliographystyle{plain}
	\title{Both cellular ATP level and ATP hydrolysis free energy determine energetically the calcium oscillation in pancreatic $\beta$-cell}
	
	\author{Yunsheng Sun} \affiliation{School of Physics, Peking University, Beijing 100871, China}
	\author{Congjian Ni} \affiliation{School of Physics, Peking University, Beijing 100871, China}
 \author{Yingda Ge} \affiliation{School of Physics, Peking University, Beijing 100871, China}
 \author{Hong Qian} \affiliation{Department of Applied Mathematics, University of Washington, Seattle, Washington 98195, USA.}
	
	\author{Qi Ouyang}
	\email{qi@pku.edu.cn}
	\affiliation{School of Physics, Peking University, Beijing 100871, China}
	\affiliation{Center for Quantitative Biology, Peking University, Beijing 100871, China}
	\author{Fangting Li}
	\email{lft@pku.edu.cn}
	\affiliation{School of Physics, Peking University, Beijing 100871, China}
	\affiliation{Center for Quantitative Biology, Peking University, Beijing 100871, China}

	\begin{abstract}
		In pancreatic $\beta$-cells, calcium oscillation signal is the core part of glucose-stimulated insulin secretion. Intracellular calcium concentration oscillates in response to the intake of glucose, which triggers the exocytosis of insulin secretory granules. ATP plays a crucial part in this process. ATP increases as the result of glucose intake, then ATP binds to ATP-sensitive $K^+$ channels ($K_{ATP}$), depolarizes the cell and triggers calcium oscillation, while the ion pumps on the cell membrane consumes the free energy form ATP hydrolysis. Based on Betram et. al. 2004 model, we construct a kinetic models to analyze the thermodynamic characteristics of this system, to reveal how the  ATP hydrolysis free energy affects the calcium oscillation. Our results suggest that bifurcation point is sensitive to both the free energy level and cellular ATP level, and the insufficient ATP energy supply would cause dysfunction of calcium oscillation.
	\end{abstract}
	\maketitle
	
	{\bf Content}
	
	I, INTRODUCTION
	
	II, MODELING
	
	III, RESULT
	
	IV, DISCUSSION
	
	\section{Introduction}
	\par
In pancreatic $\beta$-cells, calcium oscillation signal is the core part of glucose-stimulated insulin secretion. In response to the intake of glucose, the intracellular calcium concentration oscillates and couples the oscillation of membrane potential, then the calcium oscillation triggers the exocytosis of insulin secretory granules \cite{nicholls}.

All biochemical systems in living cells are open system, the steady state of the system is a non-equilibrium state that is far away from the thermodynamic equilibrium state. The Gibbs free energy of ATP hydrolysis is a typical source of energy to drive and maintain the non-equilibrium steady state. In the calcium oscillation process for the pancreatic $\beta$-cells, ATP plays an important role through ATP-sensitive $K^+$ channels ($K_{ATP}$) channel and calcium pumps on the cell membrane, where ATP influences ${K_{ATP}}$ \cite{hopkins_two_1992}and ATP hydrolysis provides energy for calcium pumps.

In the research of calcium oscillation in pancreatic $\beta$-cells, many mathematical models have taken ATP into consideration \cite{nilsson}, and ATP:ADP ratio and calcium signals has been traced with time series \cite{felix}. However, none of the works and models has analyzed the system from the thermodynamic aspect. Here we use dynamic model to analyze the thermodynamic characteristics of this system, to investigate how the cellular ATP level and ATP hydrolysis free energy affect the calcium oscillation.  Our results suggest that bifurcation depends on both the free energy level and cellular ATP level, and the insufficient ATP energy supply would cause dysfunction of calcium oscillation.
\par
According to in-vitro studies in pancreatic $\beta$-cells, there are three oscillation modes varying in the magnitude of frequency\cite{Bertram}. The fastest oscillation mode has a period of few seconds, and is generated by the interaction between plasma potential and intracellular calcium concentration. The other two are generated by oscillation of other metabolites and they display periods over 10 minutes. We only focus on the fastest mode.
	
	\section{Modeling}
	\par
	There are several excellent works to analyze the calcium oscillation in pancreatic $\beta$-cells \cite{felix}. The model of Bertram et al. involves two separated modules, glycolysis and ion transportation system, where ATP is an output of the glycolysis process into the ion transportation process\cite{Bertram}. Here, we construct a thermodynamic valid model based on the model of Bertram, we skip the step of glycolysis and treat the ATP and its free energy as direct input of the calcium oscillation system, and consider the ATP hydrolysis process as reversible reactions to reveal its thermodynamic properties. For the sake of simplification and further nonlinear dynamic analysis, we simplify Bertram's model into a two-variable dynamic model, containing the membrane potential and cytoplasm calcium concentration. We also abandon the delayed rectifier activation.

\par
\begin{figure}[H]
		\centering
		\includegraphics[width=0.7\linewidth]{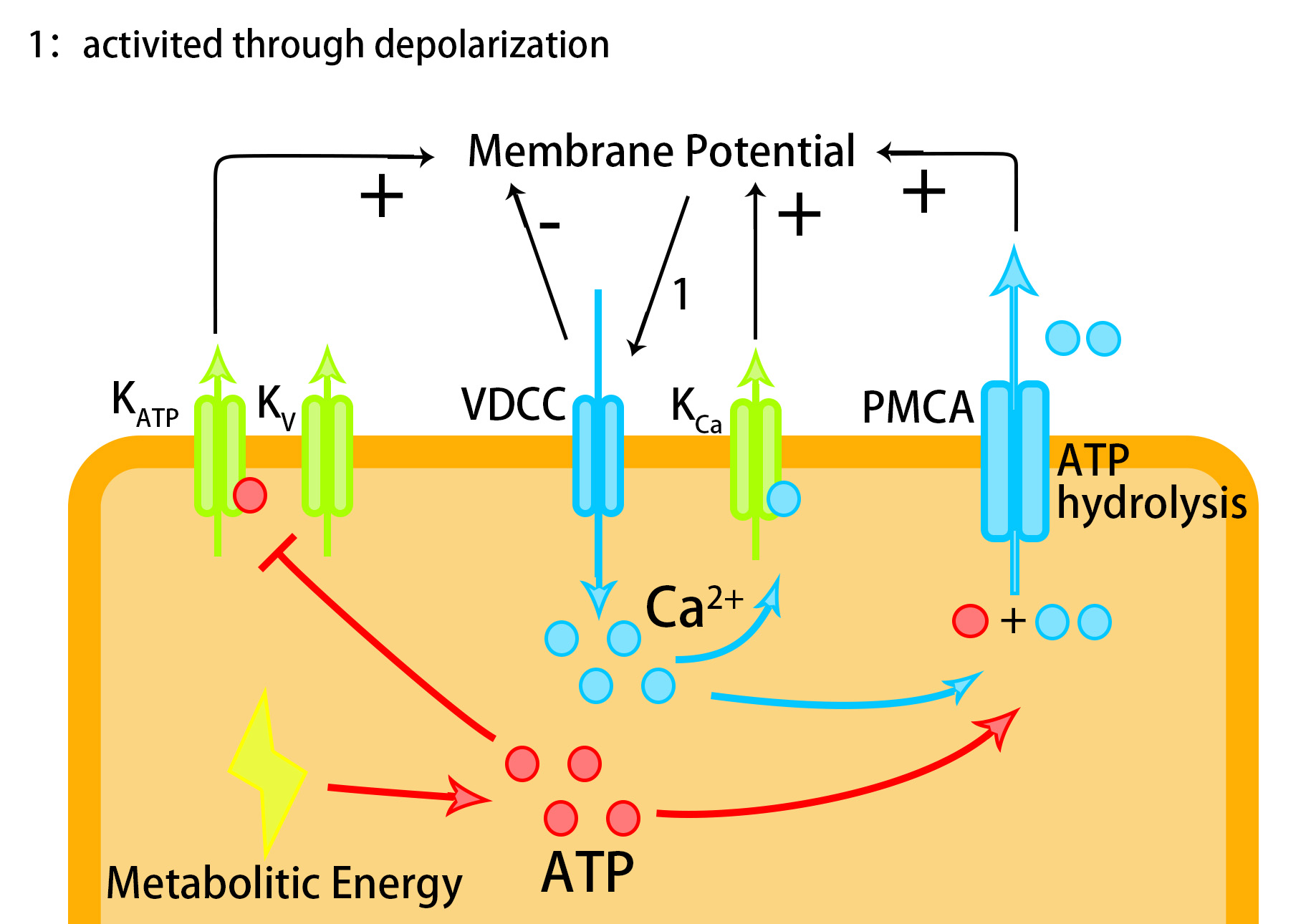}
		\caption{ Schematic calcium oscillation system in pancreatic $\beta$-cell. Ion channels such as ATP-sensitive $K^+$ channels ($K_{ATP}$), Voltage dependent calcium channel (VDCC), calcium dependent potassium channel ($K_{Ca}$), and a ligand-independent potassium channel ($K_v$) and calcium pump $PMCA$ work together to establish a feedback system involving $Ca^{2+}, K^+$, membrane potential and exhibiting oscillation. ATP is needed to trigger the oscillation through binding $K_{ATP}$, and to sustain the oscillation through providing energy for PMCA. }
		\label{fig:fig1}
	\end{figure}

Fig.\ref{fig:fig1} illustrates the schematic mechanism of fast mode calcium oscillation in pancreatic $\beta$-cells. First, as a consequence of glucose intake, ratio of ATP and ADP rises, which closes ATP-sensitive $K^+$ channels ($K_{ATP}$) \cite{hopkins_two_1992} and causes plasma membrane's depolarization. Plasma membrane depolarization activates voltage-gated calcium channel (VDCC) and introduces influx of calcium ion. Intracellular calcium will rise. Because of the increase of intracellular calcium, calcium pumps (PMCA) transport calcium ions out of cytosol at higher rate, and restore the calcium concentration in cytoplasm. The active transport consumes the free energy of ATP hydrolysis. At high calcium concentration calcium dependent potassium channel ($K_{Ca}$) will also open and restore the outflux of $K^+$, thus repolarizing the plasma membrane. As both membrane potential and [$Ca^{2+}$] are restored, a period of calcium oscillation completes. In those processes, membrane potential is affected by channels Voltage dependent calcium channel (VDCC), ATP-sensitive $K^+$ channel ($K_{ATP}$), calcium dependent potassium channel ($K_{Ca}$), and a ligand-independent potassium channel ($K_v$). Cytoplasm calcium is affected by PMCA and VDCC.
	\par
	
	\subsection{Membrane potential related channels}
	In our model, the differential equation for membrane potential V is as follows:\\
	\begin{equation}\label{potential}
	\frac{dV}{dt}=-[I_{Ca}+I_K+I_{K(Ca)}+I_{K(ATP)}]/C_m,
	\end{equation}
	
where Cm is the membrane capacitance, $I_{Ca}$ indicates calcium ion fluxes through VDCC, and $I_K$ indicates potassium ion fluxes through a ligand-independent channel $K_v$.

\par
The ionic currents are:
	\begin{equation}\label{ICa}
	I_{Ca}=g_{Ca}\cdot m_{\infty}(V)\cdot (V-V_{Ca}),
	\end{equation}
	\begin{equation}
	I_K=g_K\cdot n_{\infty}(V) \cdot (V-V_K) \label{IK},
	\end{equation}
where $g_{Ca}$ and $g_k$ are conductances for VDCC and $K_v$ respectively, $V_{Ca}$ and ${V_K}$ are Nernst potentials for calcium and potassium, $m_{\infty}$ and $n_{\infty}$ are their activation factor adopting rapid equilibrium, $v_m$ and $v_n$ are gate opening voltages.

\par
We incorporated $m_{\infty}(V)$ directly into $I_{Ca}$, and $n_{\infty}(V)$ into $I_{K}$. The steady state activation functions are:
	\begin{equation}\label{minfty}
	m_{\infty}(V)=[1+e^{(v_m-V)/s_m}]^{-1},
	\end{equation}
	\begin{equation}\label{ninfty}
	n_{\infty}(V)=[1+e^{(v_n-V)/s_n}]^{-1},
	\end{equation}
which have an increasing dependence on voltage and saturate at large positive voltages.

\par
	$I_{K(Ca)}$ indicates fluxes of $K^+$ through a kind of $Ca^{2+}$-dependent channels, and we have
		\begin{equation}\label{IKCa}
	I_{K(Ca)}=g_{K(Ca)}\cdot \omega\cdot (V-V_K).
	\end{equation}
	
	The activation factor $\omega$ is
	\begin{equation}\label{omega}
	\omega=\frac{[Ca^{2+}]^2_{cyto}}{[Ca^{2+}]^2_{cyto}+k_D^2},
	\end{equation}
where $k_D$ is the dissociation constant for $Ca^{2+}$ binding to the channel, and the $[Ca^{2+}]_{cyto}$ is the concentration of free calcium ions in cytoplasm.
	
	\par
	$I_{K(ATP)}$ indicates fluxes of $K^+$ through ATP-sensitive $K^+$ channels, we have
	\begin{equation}\label{IKATP}
	I_K(ATP)=G_{K(ATP)} \cdot \theta \cdot (V-V_K).
	\end{equation}
	The activation conductance adjusts instantaneously to the concentrations of ADP and ATP (but not free energy). The form of the open probability $\theta$ of ATP-sensitive $K^+$ channels ($K_{ATP}$) is described as:
	\begin{equation}\label{theta}
	 \theta=\frac{0.08(1+2[MgADP^-]/k_{dd})+0.89([MgADP^-]/k_{dd})^2}{(1+[MgADP^-]/k_{dd})^2(1+[ADP^{3-}]/k_{td}+[ATP^{4-}]/k_{tt})},
	\end{equation}
where $[MgADP^-]=0.165[ADP], [ADP^{3-}]=0.135[ADP], [ATP^{4-}]=0.05[ATP]$. The above equation of $\theta$ on $K_{ATP}$ is based on the results of Magnus and Keizer (1998a)\cite{magnus} that fitting the experimental data of the open probability of $K_{ATP}$\cite{hopkins_two_1992}. More evidence about the function of ATP hydrolysis and ATP level in $K_{ATP}$ \cite{zingman_signaling_2001} are discussed in section III-C.
	\subsection{Calcium pumps as active transport channels}
	\par
	Active transport of cytosolic calcium concentration is mainly through the transportation by ATPase PMCA. It works as a calcium pump. When $[Ca^{2+}]_{cyto}$ rises, the calcium pumps can transport calcium ions into high concentration calcium pools (either ER or plasma) by consuming the free energy from ATP hydrolysis. In our model, we only considered the calcium pools of the plasma, and the calcium pumps PMCA transport $Ca^{2+}$ through the cell membrane to the plasma.
Because two $H^+$ ions cross the membrane through ATPases in the opposite direction as each $Ca^{2+}$ ion is transported\cite{levy}, the membrane potential doesn't affect this ion transportation. By considering ATP hydrolysis process as reversible, we obtain a thermodynamical available model. Based on Inesi's work\cite{Inesi}, we simplify the calcium pumps into a two-step reaction:
	\begin{equation}
	 2Ca_{cyto}^{2+}+E_1+ATP\autorightleftharpoons{$k_1$}{$k_{-1}$}E_2+ADP,
	\end{equation}
	\begin{equation}
	 E_2\autorightleftharpoons{$k_2$}{$k_{-2}$}2Ca^{2+}_{pla}+E_1+Pi.
	\end{equation}

\par
Then we have the kinetic equations for calcium pumps PMCA:
	\begin{equation}
	J_{PMCA}=2k_{-1}[E_2][ADP]-2k_1[Ca^{2+}]_{cyto}^2[E_1][ATP],
	\end{equation}
	\begin{equation}
	 \frac{d[E_1]}{dt}=k_{-1}[E_2][ADP]-k_1[Ca^{2+}]_{cyto}^2[E_1][ATP]-k_{-2}[Ca^{2+}]_{pla}^2[E_1][Pi]+k_{2}[E_2],
	\end{equation}
	\begin{equation}
	-J_{PMCA}=2k_{2}[E_2]-2k_{-2}[Ca^{2+}]_{pla}^2[E_1][Pi],
	\end{equation}
	\begin{equation}
	 \frac{d[E_2]}{dt}=-k_{2}[E_2]+k_{-2}[Ca^{2+}]_{pla}^2[E_1][Pi]-k_{-1}[E_2][ADP]+k_1[Ca^{2+}]_{cyto}^2[E_1][ATP],
	\end{equation}
	\par
 where the $[Ca^{2+}]_{cyto}$ is free calcium ions in cytoplasm, $J_{PMCA}$ is the rate for $Ca^{2+}$ passing through PMCA; $[E_1]$ and $[E_2]$ represent the unbinding and binding form of ATPases respectively, and $[E_1]+[E_2]=[E_T]$, a constant which is the total amount of ATPases.

 \par
 If we adopt quasi-steady assumption, where $\frac{d[E_1]}{dt}=\frac{d[E_2]}{dt}=0$, the net transportation rate through ATPases can be derived:
	\begin{equation}\label{JPMCA}
	 J_{PMCA}=k_{pmca}\frac{k_1k_2[ATP][Ca^{2+}]_{cyto}^2-k_{-1}k_{-2}[ADP][Pi][Ca^{2+}]_{pla}^2}{k_{-1}[ADP]+k_1[Ca^{2+}]_{cyto}^2[ATP]+k_{-2}[Ca^{2+}]_{pla}^2[Pi]+k_{2}}
	\end{equation}
	\par
	
$k_{pmca}$ substitutes $2[E_T]$ for convenience. Since the calcium pool in plasma is large, we consider the $[Ca^{2+}]_{pla}$ as a constant. Also although there is consumption of ATP, its affect on ATP concentration can be ignored. According to thermodynamic law and non-equilibrium steady state theory\cite{qian}, there is additional relationship that $\frac{k_1 k_2}{k_{-1} k_{-2}}=K_{eq}=4.9 \times 10^{11}$.

We have the equation for $[Ca^{2+}]_{cyto}$ as:
	
	\begin{equation}\label{Cacyto}
	\frac{d[Ca^{2+}]_{cyto}}{dt}=(-\alpha I_{Ca}-J_{PMCA})f_{cyt}.
	\end{equation}
\par

Considering only a fraction of $Ca^{2+}$ that enters cytosol becomes free form, we use $f_{cyt}=0.01$ to denote the ratio\cite{Bertram}.
	\par
	
Thus, we obtain the thermodynamic valid 2-variable model to depict the fast mode of calcium oscillation in pancreatic $\beta$-cells, they are Eq.\ref{potential} and Eq.\ref{Cacyto}. We hope this simplified model help us to reveal the function of free energy caused by ATP hydrolysis in calcium oscillation process.

	\par
	\subsection{[ATP] and $\gamma$ affect the membrane potential and  $[Ca^{2+}]_{cyto}$ }
	The reaction of calcium pumps incorporate ATP hydrolysis, we analyze the role of ATP hydrolysis free energy in the calcium pumps transportation. Mean while, we analyze the role of cellular ATP level and the ATP hydrolysis free energy in the ATP-sensitive $K^+$ channels($K_{ATP}$).
\par
The free energy of ATP hydrolysis is defined by laws of thermodynamics as
	
	\begin{equation}\label{deltag}
	\Delta G_{ATP}=\Delta G_{ATP}^0+k_BTln\frac{[ADP][Pi]}{ATP}.
	\end{equation}
	
	For convenience, we introduce
	\begin{equation}\label{gamma}
	\gamma=e^{-\Delta G_{ATP}/(k_BT)}=K_{eq}\frac{[ATP]}{[ADP][Pi]},
	\end{equation}
	
Where $K_{eq}$ is the equilibrium constant of ATP hydrolysis under standard condition ($K_{eq}=4.9 \times 10^{11}\mu M$, according to avalible data\cite{qian}).

$\gamma$ is proportional to the reaction quotient of ATP hydrolysis, to substitute $\Delta G_{ATP}$. $\gamma$ determines the direction of the hydrolysis reaction. When the biochemical reactions are at thermodynamic equilibrium state, $\gamma=1$. Under the physiological condition, $\gamma \approx 10^{10}$,  the concentration for ATP and ADP are $10^3 \mu M$ and $10^2 \mu M$ respectively, [Pi] is about 1 mM\cite{schell}. In our simulation and analysis, we assume that [Pi] is constant (1 mM), so any two of [ATP], [ADP] and $\gamma$ are independent variables, and we set ATP and $\gamma$ as independent variables in the following part.
	\par
	In terms of [ATP] and $\gamma$, $J_{PMCA}$ in Eq.\ref{JPMCA} can be rewritten as
	\begin{equation}\label{JPMCA2}
	 J_{PMCA}=k_{pmca}\frac{k_1k_2[ATP][Ca^{2+}]_{cyto}^2(1-\frac{1}{ \gamma }\frac{[Ca^{2+}]_{pla}^2}{[Ca^{2+}]_{cyto}^2} )}{k_{-1}K_{eq}\frac{[ATP]}{\gamma [Pi]} +k_1[Ca^{2+}]_{cyto}^2[ATP]+k_{-2}[Ca^{2+}]_{pla}^2[Pi]+k_{2}}
	\end{equation}
	\par
	The reaction direction is determined by the sign of numerator, which is determined by $\gamma$. If $\gamma$ is too low, reaction direction would be reversed and active transportation can not be reached, so there should be no calcium oscillation.
	\par
	Similarly, the open probability of $K_{ATP}$, $\theta$, (Eq.\ref{theta}) can be rewritten in terms of [ATP] and $\gamma$:
	\begin{equation}\label{theta2}
	\theta=\frac{0.08+0.026 \frac{1}{k_{dd}}\frac{K_{eq}[ATP]}{\gamma [Pi]}+0.024(\frac{1}{k_{dd}}\frac{K_{eq}[ATP]}{\gamma [Pi]})^2}{(1+0.165\frac{1}{k_{dd}}\frac{K_{eq}[ATP]}{\gamma [Pi]})^2(1+0.135\frac{1}{k_{td}}\frac{K_{eq}[ATP]}{\gamma [Pi]}+0.05[ATP]/k_{tt})}
	\end{equation}

\par
More analysis and simulation results about how the $[ATP]$ and $\gamma$ influence the $J_{PMCA}$ and $\theta$ is shown in section III-C ``Thermodynamic analysis of ATP-sensitive $K^+$ channels ($K_{ATP}$) and calcium pump".
	
	\section{Result}

	\subsection{Oscillation border in parameter space is determined by both [ATP] and $\gamma$ (or [ADP])}
	
\begin{figure}[H]
		\centering
		\includegraphics[width=\linewidth]{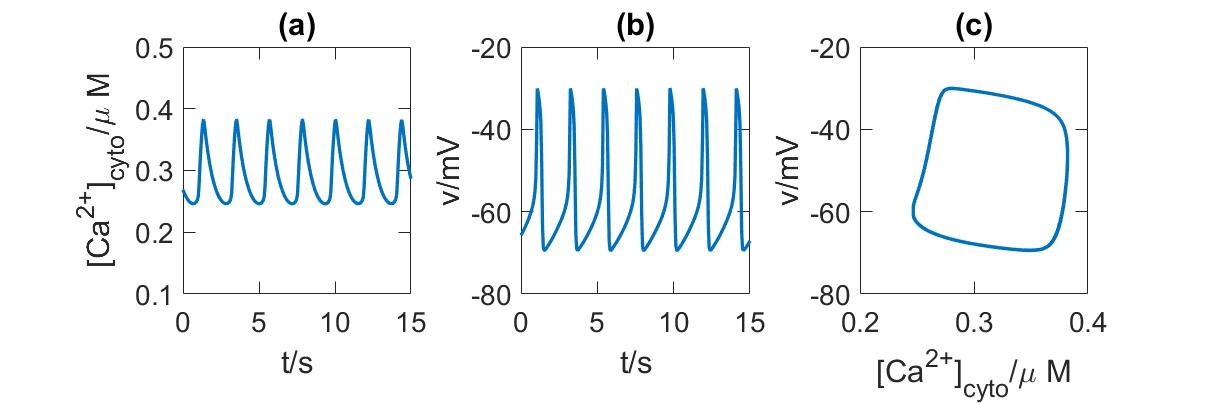}
		\caption{The fast calcium oscillation mode in our model under $[ATP]=1400\mu$M,$\gamma=10^{10}$, the frequency of oscillation is 0.46Hz. (a) The oscillation of $[Ca^{2+}]_{cyto}$ with amplitude=0.13$\mu M$. (b) The oscillation of membrane voltage with amplitude=39mV. (c) The phase portrait of $[Ca^{2+}]_{cyto}-v $.}
		\label{fig:single}
	\end{figure}

We assume that the concentrations of each component in ATP hydrolysis stay at some certain constant values, so our 2-variable model aims to explain the fast calcium oscillation mode. Because the concentration of [Pi] is relative high (1000$\mu M$) and stable\cite{qian}, we set [Pi] as constant, and utilize the concentration of ATP, [ATP], and $\gamma$ as key parameters to describe ATP hydrolysis process. The parameters are listed in Table 1. The simulation of time series of our model is calculated by ode15s function of MATLAB. When [ATP]=1400 $\mu M$, $\gamma=10^{10}$, the calcium oscillation is shown in Fig. \ref{fig:single}, where the time period between pulses is 2.2 seconds and the amplitude of calcium oscillation ($[Ca^{2+}]_{cyto}$) is 0.13$\mu M$, which matches the fast oscillation in experiment \cite{Liu}. However, when the concentration of ATP or the level of and $\gamma$ is not high enough, such as [ATP]=900 $\mu M$ and $\gamma=10^{10}$, the system doesn't oscillate.
\par	
Thus [ATP] or $\gamma$ can determine whether the system is in oscillatory state. Under varying [ATP] or $\gamma$ the system can become unstable and start oscillating, which is the non-linear physics term "bifurcation".

	\par
	We first quantified the function of ATP hydrolysis in the calcium oscillation by our 2-variable model Eq.\ref{potential} and \ref{Cacyto}. In the Eq.\ref{potential} of membrane potential $V$, the change of [ATP] and $\gamma$ would affect the ATP-sensitive $K^+$ channels, $K_{ATP}$. When more ATP binds to $K_{ATP}$, $K_{ATP}$ would be blocked, this results in depolarization of $\beta$-cells and triggers the calcium oscillation. $\gamma$ can also affect $K_{ATP}$ channel. In the Eq. \ref{Cacyto}, ATPases consume energy to transport calcium against concentration gradient, $\gamma$ significantly influence the ATPases like PMCA. Through these two channels [ATP] and $\gamma$ is affecting calcium oscillation. More quantitative analysis will be provided in the following part.
	
	\par
\begin{figure}[H]
	\centering
	\includegraphics[width=\linewidth]{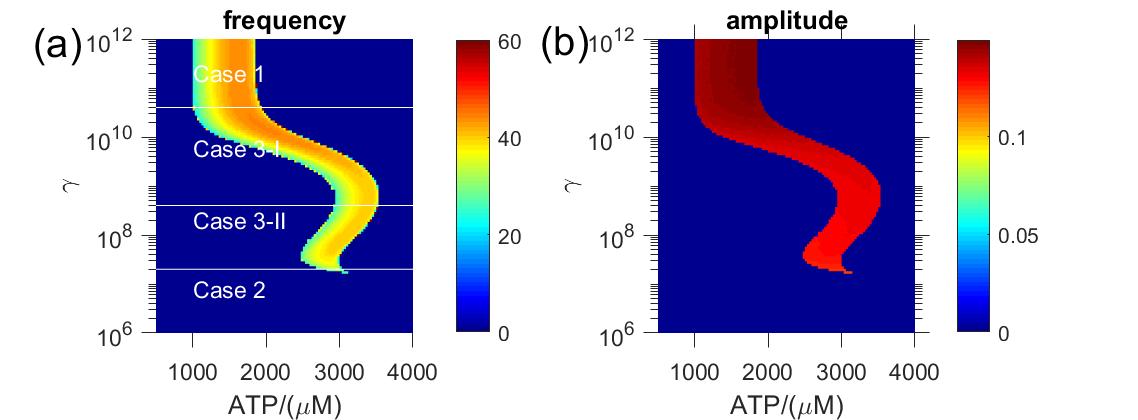}
	\caption{Calcium signals in steady state (in blue) or oscillation (in green, yellow or red) under different values of [ATP] and $\gamma$ . Each point stands for a time series under a certain [ATP] and $\gamma$ value. (a) The frequencies of the calcium oscillations, where the base frequency is extracted by fast Fourier transformation (FFT). (b) The amplitudes of calcium signals, which is defined as the peak-peak values in each time series with stable oscillation.  Based on the behavior of bifurcation border, we divide the range of $\gamma$ into 4 regions: Case 1, Case 2, Case 3-I and II.}
	\label{fig:twodimension}
\end{figure}

	We mapped the amplitudes and frequencies of those oscillations under different [ATP] and $\gamma$. It is shown in Fig. \ref{fig:twodimension}. A colored region indicates oscillation exist in such [ATP] and $\gamma$ values, while region in mono dark blue means no oscillation. Border of the region is where the transition between stable and oscillating state (the transition is called bifurcation) happens. Colors in oscillating region represent frequency and amplitude in the two sub-figures respectively. We find that the amplitudes almost remain constant while the frequencies are much lower near bifurcation.
	\par
	First we need to explain the independent role of [ATP] and $\gamma$ respectively. For varying [ATP], the system oscillates within a certain range of it. Suppose there is a increase in [ATP] with fixed $\gamma$ value. When [ATP] first reaches a certain value, $K_{ATP}$ channel will be closed and oscillation starts, as indicated by Fig \ref{fig:twodimension}. That is a bifurcation. Where [ATP] becomes too high, there is no oscillation because membrane potential can not be restored then.
	\par
	Then we investigated the role of $\gamma$. Under different values of $\gamma$, the bifurcation points vary. In regard to the property of bifurcation points' response to $\gamma$, the figure can be divided into 4 regions: Case 1, Case 2, Case 3-I and II.
	\par
	When $\gamma>\gamma_1\approx 10^{11}$, which is Case 1, $\gamma$ yields no impact on bifurcation. Because when $\gamma$ is in this high level, the reaction are nearly irreversible. We consider the region of higher $\gamma$ to be trivial extrapolation. It should be noticed that Case 1 is the region where ATPases' reaction is approximately irreversible, as previous Bertram's model. In our model this region is above the physiology normal value of $\gamma\approx 10^{10}$ for magnitudes. In fact, under the physiology value of $\gamma\approx 10^{10}$, the reactions of ATPases shouldn't be treated as irreversible ones.
	\par
	When $\gamma<\gamma_2\approx 10^{7}$, which is the Case 2, our model shows that the calcium oscillation can not be generated. The reason is that PMCA lacks the energy to transport calcium against the ion concentration gradient. This demonstrates the idea that energy consumption is necessary for non-equilibrium steady state \cite{qian}.
	\par
	The nontrivial result is the Case 3-I and II with $\gamma_2<\gamma<\gamma_1$, where the bifurcation border bends. [ATP] and $\gamma$ affect the bifurcation through the channels of $K_{ATP}$ and PMCA. Here we utilized the non-linear dynamical method of null-lines on the phase plane to analyze this problem, as is shown in the following subsection.
	\par
	In Supplemental Information I, we also discussed a 4-variable model considering endoplasmic reticulum and a delayed rectifier activation through $K_V$ channel, and obtained similar results as Fig. \ref{fig:twodimension}.
	
	\subsection{Analysis of Calcium Oscillation on Phase Plane}
	Based on the simplified 2-variable model, we investigate and analyze the fast mode of calcium oscillation in pancreatic $\beta$-cells on the phase plane of membrane potential $V$ and calcium concentration in cytoplasm $[Ca^{2+}_{cyto}]$.
	\par
\begin{figure}[H]
		\centering
		\includegraphics[width=\linewidth]{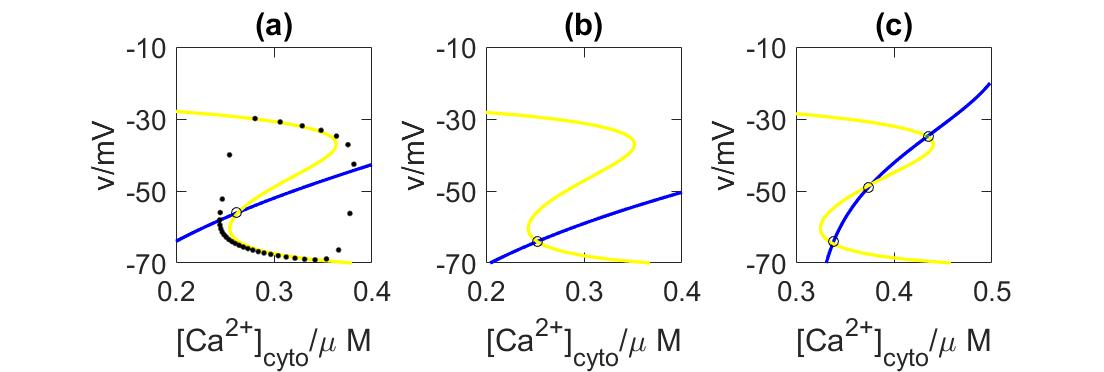}
		\caption{Phase plane of the calcium oscillation system. Blue line: $\frac{d[Ca^{2+}]}{dt}=0$ nullcline. Yellow line: $\frac{dV}{dt}=0$ nullcline. Dot line: stable trajectory of time sequence simulation. 'O' point: fixed point. (a) Under [ATP]=1600$\mu M$ and $\gamma=10^{10}$, the two nullclines yield a unstable fixed point, and a limit cycle emerges. This is a "oscillation caused by hysteresis", and the bifurcation is a Hopf one.(b) Under $[ATP]=1000\mu M$ and $\gamma=10^{10}$, the decreasing [ATP] makes the fixed point stable, and there is no limit cycle. (c) Under low $\gamma$ ($\gamma=10^{7}, [ATP]=4400\mu M$), the two nullclines intersect at unstable fixed point, there are two additional stable fixed point which eliminate the limit cycle. The system is globally stable.}
		\label{fig:phasemap}
	\end{figure}

	In Fig.\ref{fig:phasemap}, we illustrate the bifurcation of potential $V$ and $[Ca^{2+}]_{cyto}$ with different [ATP]. At low ATP concentration (Fig.\ref{fig:phasemap} (b)), the V-nullcline ($\frac{dV}{dt}=0$) and the $[Ca^{2+}]_{cyto}$ - nullcline ($\frac{d[Ca^{2+}]_{cyto}}{dt}=0$) intersect and yield a stable fixed point, a stable spiral in fact. With the raise of [ATP], the nullclines move and the fixed point becomes unstable (Fig.\ref{fig:phasemap} (a)). Nearby trajectories are all driven to a limit cycle, namely a bifurcation. It is in fact a subcritical Hopf bifurcation. The limit cycle trajectory are displayed as discrete points with same time interval. Higher density of dots represents lower speed of evolution. The trajectory gets blocked near the fixed point, since the derivative of V and $[Ca^{2+}]_{cyto}$ are lower there. In Fig. \ref{fig:twodimension}, the oscillation frequency near bifurcation border is lower. Around bifurcation border the fixed point is near to the trajectory, and this can explain the frequency change.
	\par
\begin{figure}[H]
		\centering
		\includegraphics[width=\linewidth]{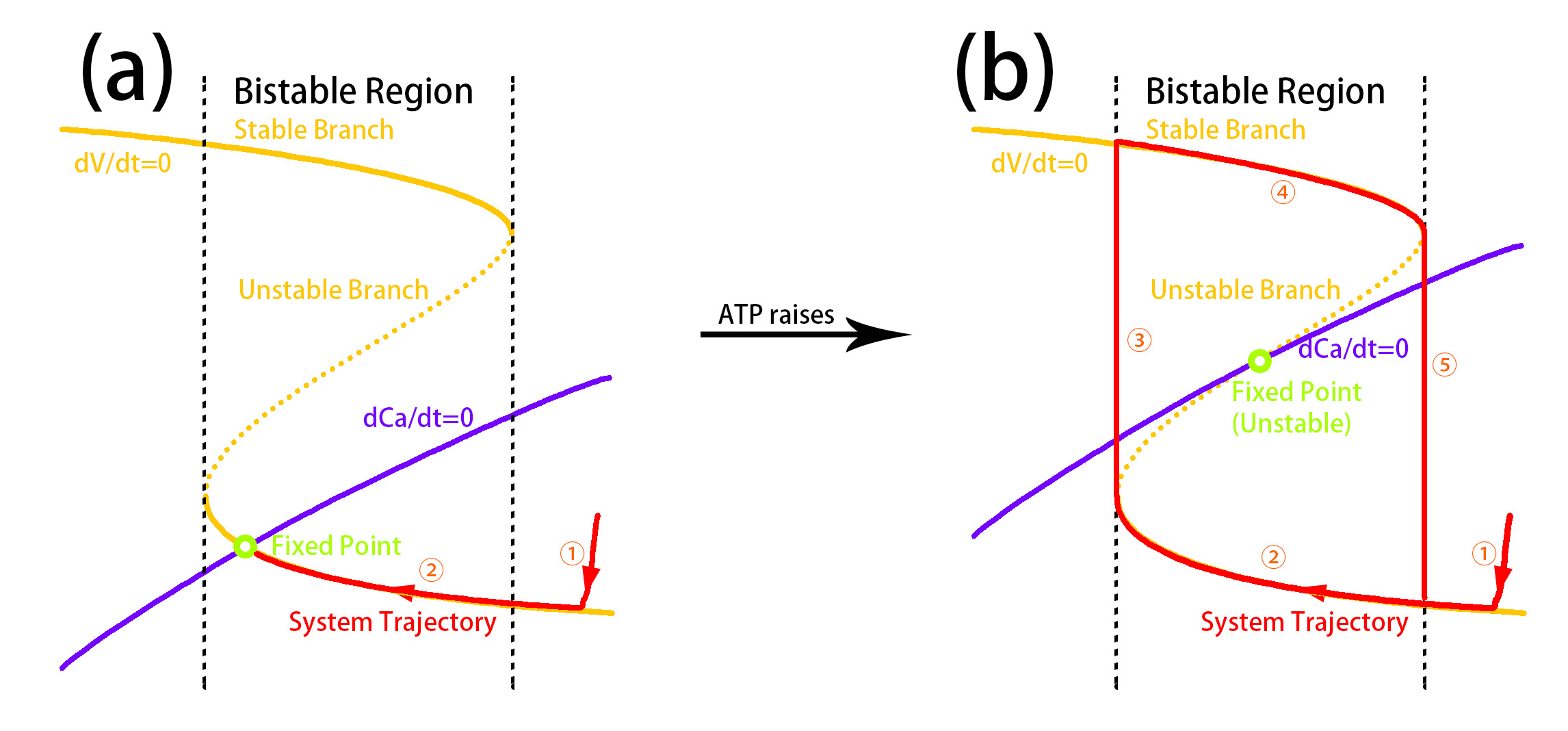}
		\caption{Illustration of bifurcation in calcium oscillation system on phaseplane. Yellow: V-nullcline (dV/dt=0). Solid: stable branch. Dotted: unstable branch. Blue: $[Ca^{2+}]$-nullcline. Red: Trajectory of the system. Green dot: fixed point. (a): The system approaches V-nullcline first, it will then evolve towards the Ca-nullcline. The fixed point is stable and system would rest at that state.(b): With higher [ATP], nullclines move and the fixed point becomes unstable. The system falls off the stable branch into another and reaches an cycling state.}
		\label{fig:phaseplane}
	\end{figure}

	Fig. \ref{fig:phaseplane} explains how the stability of fixed points is influenced by the displace of nullclines. In our model, the evolve rate of V is often much more larger than that of $[Ca^{2+}]_{cyto}$. The system would first reach the stable state of V under a certain $[Ca^{2+}]_{cyto}$, and the region where V-nullcline folds is a bistable region. After the system approaches the V-nullcline, it will then evolve towards the $[Ca^{2+}]_{cyto}$ - nullcline. In the case of the left figure, the system reaches the stable fixed point by evolving on the stable branch of V-nullcline. However in the right figure, the system can't reach the fixed point, and it falls off the stable branch into another. Thus the system reaches an oscillating state. Such bifurcation has a term called 'oscillation caused by hysteresis'. This also explains why near the bifurcation border, the fixed point is close to the oscillation trajectory. If [ATP] increases further, another bifurcation occurs and the system would be stable again. These explains why varing [ATP] results in two bifurcation branches in Fig. \ref{fig:twodimension}.
	
	\subsection{Thermodynamic analysis of ATP-sensitive $K^+$ channels ($K_{ATP}$) and calcium pump}
\begin{figure}
	\centering
	\includegraphics[width=0.7\linewidth]{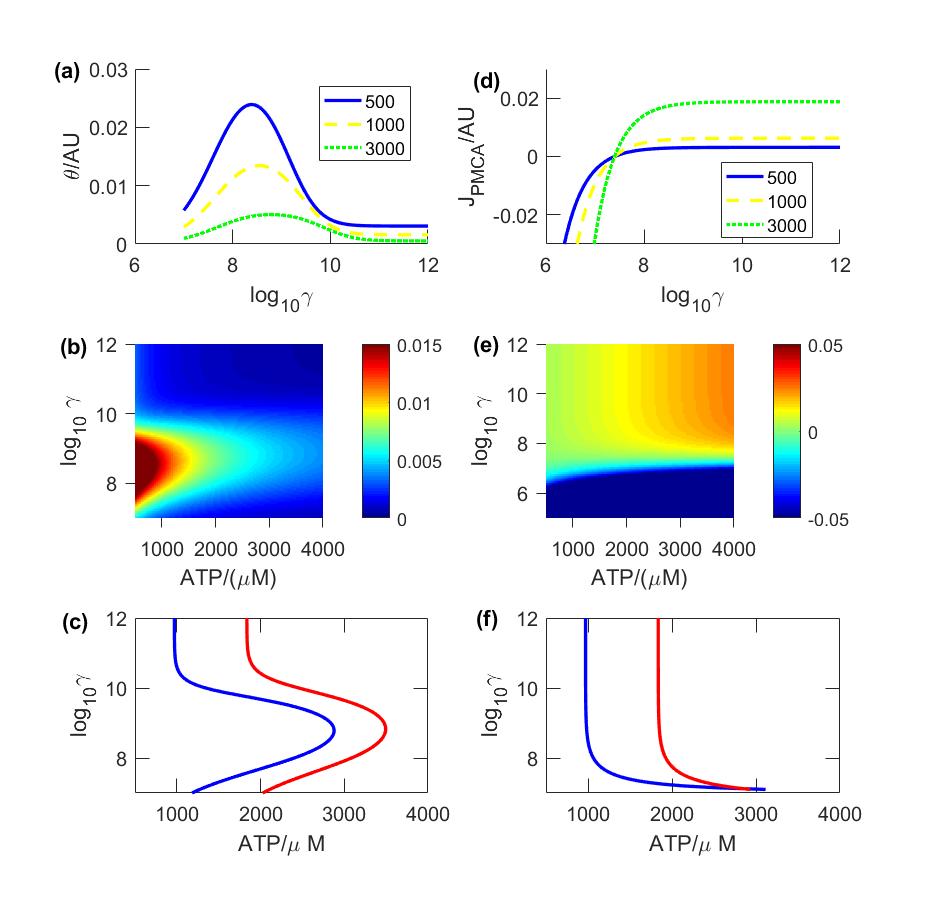}
	\caption{Thermodynamic analysis of ATP-sensitive $K^+$ channels ($K_{ATP}$) and calcium pumps. (a) $\gamma$ affects the open probability $\theta$ (in Eq. \ref{theta}) of $K_{ATP}$ . The open probability would be largely enhanced when $\gamma<\gamma_1\approx 10^{11}$. (b) The value of   $\theta$ under different [ATP] and $\gamma$. The significate change in $\gamma<\gamma_1\approx 10^{11}$ leads to the bend shape of bifurcation border in this region. (c) Bifurcation border with only $K_{ATP}$ related to $\gamma$ (PMCA is modified). It can be seen that the bend in bifurcation border is caused by $K_{ATP}$. (d) $J_{PMCA}$'s dependance on $\gamma$. $[Ca^{2+}]_{cyto}$ is fixed at 0.2$\mu M$. At low $\gamma$ level ($\gamma<\gamma_3 \approx 10^{8}$) the function of PMCA is significantly affected. The pump would be reversed for $\gamma<10^{7}$. (d) $J_{PMCA}$ under different [ATP] and $\gamma$. (f) Bifurcation border with only PMCA related to $\gamma$ ($K_{ATP}$ is modified.). It proves that oscillation cut off in the low $\gamma$ region is caused by the reversibility of PMCA.}
	\label{fig:channel}
\end{figure}

	In Case 3-I and Case 3-II, with $\gamma_2<\gamma<\gamma_1$ bifurcation border bends with varying $\gamma$. This means bifurcation points are affected by $\gamma$. That dependance is a result of enhanced open probability $\theta$ (as defined in Eq.\ref{theta}) of $K_{ATP}$ within certain range of $\gamma$. Relationship between open probability $\theta$ of $K_{ATP}$ and $\gamma$ is shown in Fig. \ref{fig:channel} (a). Around physiology level of $\gamma$ ($\gamma \approx 10^{10}$), the open probability of $K_{ATP}$ is largely enhanced.  That is further demonstrated by Fig. \ref{fig:channel} (b), which is the open probability in plane of ATP and $\gamma$. In this way, high [ATP] level is necessary to restore it in such region. In the phase plane, it means high [ATP] is needed to restore the position of V-nullcline. In fact $[Ca]^{2+}_{cyto}$-nullcline is hardly affected in case 3. This leads to the shifts of bifurcation border.
	\par
	In our model, PMCA and $K_{ATP}$ channels are the only two that depends on $\gamma$ value. To investigate the function of $K_{ATP}$, in Fig. \ref{fig:channel} (c) we modified the PMCA's equation to make the system related to $\gamma$ only through $K_{ATP}$. How is that achieved is explained later. Despite the modification, the bend of bifurcation border still remains. In fact, the high open probability in the bending region requires higher ATP to close $K_{ATP}$ channel and trigger the oscillation. This proves $K_{ATP}$ is the cause of border bending.
	\par
	The bend of bifurcation borders in both Case 3 I (towards higher [ATP]) and Case 3 II (towards lower [ATP]) correlate with the change in $K_{ATP}$ open probability. The function of each case is different, though. In case 3 I, lower energy would cause difficulty in generating oscillation, while in case 3 II the effect is contrary. It should be noticed that case 3 I is around physiology level of $\gamma$, and would be the common case.
	\par
	The above results about the $\gamma$ region in case 3 is supported by the experiment observations and parameter sensitivity analysis of our model (Appendix-II), so it is not a special results of our model based on a special set of parameters. In above, we find that the bend of bifurcation border directly depends on open probability $\theta$ of $K_{ATP}$. The equation of $\theta$ in our model is based on the experimental data\cite{hopkins_two_1992}, where they demonstrated that the change of open probability exists in $\gamma_2<\gamma<\gamma_1$.
	\par
Through the mechanism how the $\gamma$ affects the open probability $\theta$ of $K_{ATP}$ is not clear, some researches have revealed the role of ATP hydrolysis plays in $K_{ATP}$. $K_{ATP}$ has been proved to possess ATPase activity\cite{matsuo1999atp}. A theory that $K_{ATP}$ need ATP hydrolysis to close pore is established through experiment on trapping transition states of SUR (a subunit of $K_{ATP}$, a sulfonylurea receptor which engages in ATP hydrolysis and pore opening) during ATP hydrolysis \cite{zingman_signaling_2001}. This mechanism might explain the unique $\gamma$ dependance.

In this way, the $\gamma$ interval where bifurcation border bends is determined by experimental data.

Furthermore, we have analyzed the parameter sensitivity numerically (Appendix B) to demonstrate numerically that the bend of border is robust to the change of other parameters.

	\par
	Furthermore, we investigate how PMCA, the calcium pump that exports calcium ions, consumes free energy to maintain normal cellular function in calcium oscillation. With $\gamma>\gamma_3\approx 10^8$, PMCA is hardly affected. However impact on PMCA would be critical when $\gamma$ is under $\gamma_3$. The relationship between $J_{PMCA}$, which is the flux through PMCA, and $\gamma$ is shown in Fig. \ref{fig:channel} (d). When $\gamma$ is relatively high, $\gamma>\gamma_1\approx 10^{10}$, the reaction of ATPase is nearly irreversible, $J_{PMCA}$ would saturate to the change of $\gamma$. With low $\gamma$ level ($\gamma<\gamma_3\approx 10^8$) the function of PMCA is significantly affected and reversible, and is different from the irreversible case. There is a minimal $\gamma_4\approx 10^7$, under which oscillation no longer exist.
	\par
	We can derive a theoretical lower bound of the minimal $\gamma_4$ value. The necessary condition is that $J_{PMCA}>0$, and we can get $\gamma>\frac{[Ca^{2+}]^2_{pla}}{[Ca^{2+}]^2_{cyto}}$. Since the magnitude of $[Ca^{2+}]^2_{pla}$ and $[Ca^{2+}]^2_{cyto}$ are relatively stable and well supported\cite{qian, Bertram, Liu}, the lower bound $\gamma_{lb}\approx 10^7$ is valid. The lower bound has a good indication of the real minimal $\gamma$ value $\gamma_4$.
	\par
	To investigate the function of PMCA, we set the $\gamma$ related term in $K_{ATP}$ equation to be zero ($1/k_{dd}=1/k_{td}=0$), and the ATP sensitive $K^+$ channel is not affected by $\gamma$. The ATPase's dependance on $\gamma$ still remains. The distinctive shape of bifurcation border is lost (Fig. \ref{fig:channel} (f)) because the reason explained above, but we still have the cut off in low $\gamma$ region. That is because the ATPases don't have enough energy to support the oscillation. It is consistent with the result in the activation process of yeast S phase checkpoint, where there is a positive feedback loop for the phosphorylation of key kinase Rad53 \cite{jin_non-equilibrium_2018}.
	\par
	PMCA reaction that doesn't depend on $\gamma$ is achieved by setting $\gamma$ related terms in equation \ref{JPMCA} to be zero. ([Pi] terms are considered not to be related with $\gamma$, since we have already assumed [Pi] to be constant in previous section.) In this way Eq.\ref{JPMCA} happens to transform into Michaelis-Menten form, and the reaction is irreversible:
	\begin{equation}
	 J_{PMCA}=k_{pmca}\frac{k_1k_2[ATP][Ca^{2+}]_{cyto}^2}{k_1[Ca^{2+}]_{cyto}^2[ATP]+k_{-2}[Ca^{2+}]_{pla}^2[Pi]+k_{2}}.
	\end{equation}
	The bifurcation border in high $\gamma$ level is similar, but oscillation is not cut off by low $\gamma$ in irreversible one, as is shown in Fig \ref{fig:channel} (c). This proves the necessity to consider reverse reaction as the $\gamma$ related PMCA in our model.
	\par
	On the phase plane, it can be seen that low $\gamma$ add stable fixed points to the limit cycle. In Fig.\ref{fig:phasemap} (c), under $\gamma$ as low as $10^{7}$, $[Ca]^{2+}_{cyto}$-nullcline shifts dramatically. Two other stable fixed points emerge as the nullclines intersect at unstable branch. In this way, the system is globally stable.

	\subsection{The glucose-stimulated calcium oscillation and insulin secretion}
	
\begin{figure}[H]
	\centering
	\includegraphics[width=0.7\linewidth]{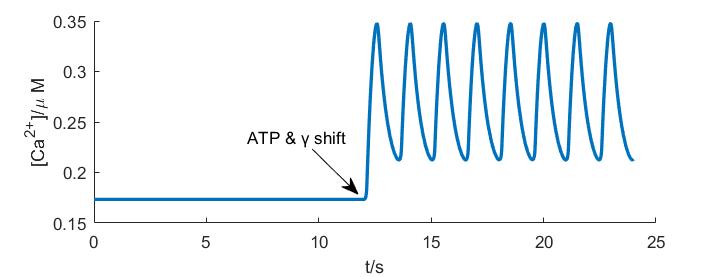}
	\caption{The simulated calcium concentration under a shift in ATP/$\gamma$ value. The initial [ATP]=1.80mM, $\gamma$=9.40, and the state is stable. After addition of glucose ATP shifts to 2.03mM and $\gamma$=9.75, oscillation begins \cite{detimary_interplay_1998}.}
	\label{fig:oscpattern2}
\end{figure}

	We know that in the normal physiology state the glucose intake will increase the cellular ATP level, we simulation and analyze the phenomenon in Fig. \ref{fig:twodimension} and Fig.\ref{fig:oscpattern2}. In the glucose-stimulated insulin secretion process, an in vitro work showed that when the level of glucose increases from 3mM to 10mM, the ratio of ATP:ADP raises from 5.1 to 11.6; $\gamma$ is from 9.40 to 9.75; and ATP concentration would raise 13\%, while the absolute ATP concentration is not mentioned\cite{detimary_interplay_1998}. Then in Fig.\ref{fig:oscpattern2}, we assume that there is a instant shift of [ATP] and $\gamma$, our simulation result suggests that the system evolves form the stable region to the oscillating region. This provides a description of calcium oscillation induced by glucose intake. In real organism system the transition of the parameters should be continuous and require time, but we just focus on the stable behavior of the system.
	\par
	If $\Delta G$ becomes lower due to some reasons, in some regions the bifurcation border moves sensitively to the direction of higher [ATP] (Case 3-I), or even vanishes (Case 2), making it harder to generate the oscillation, which indicates that around such parameters $\gamma$ dominates the bifurcation process. Case 3-I where the border bend sharply under variance of $\gamma$ is not a rare case in parameter space, and it is close to physiology level where $\gamma \approx 10^{10}$ and $[ATP] \approx 1000 \mu M$. It indicates that deficiency in metabolism and energy supply may cause dysfunction of calcium oscillation. Case 2 where the oscillation behavior is cut off by low $\gamma$ is a definite result of our model since oscillation depends on energy supply.
	\par
	Moreover, it is observed that the oscillation disappears when ATP gets too high\cite{li2013oscillations}. In our model, it can be explained as ATP passes through the second bifurcation border.
	\par
	Therefore it is naturally to presume that an increase in [ATP] or $\gamma$ results in the calcium oscillation, which is exactly what glucose-stimulated insulin secretion tells us. Furthermore, a shift in the oscillation border as well as weaker cell function in raising its ATP concentration are among the multiple reasons that cause the abnormality of $\beta$-cell.
	
	\section{Discussion}
	\par
	Living cells are thermodynamically considered as open systems that operate far from the equilibrium state. In calcium oscillation, the system is in non-equilibrium, and active energy consumption is needed to sustain it.\cite{qian}. In this work, we construct a thermodynamic model containing the reversible reaction of calcium pumps (PMCA) driven by ATP hydrolysis. When ATP hydrolysis free energy level is extremely high, we can treat the calcium pumps as irreversible. The calcium oscillation is only determined by the level of ATP. However, when the free energy is low, especially when it is unable to support the calcium pumps, the effect is prominent. Oscillation can not be sustained. The interesting case occurs in the middle part of ATP hydrolysis free energy, where both of the ATP level and ATP hydrolysis free energy determine the calcium oscillation. Thus we demonstrate the necessary role of energy consumption in living system.
	\par
	The main results highlighted from our toy model is independent from the exact form and number of variables in the models. The calcium pump consumes energy and sustains oscillation, as well as the ATP sensitive $K^+$ channel play important role in the system. These two channels determine calcium oscillation's dependence on $\gamma$. In ATPases function, the role of free energy is frequently mentioned like in Na/K pumps\cite{tran}. We further investigate this by putting the ATPases in a system, and investigate the effect on the system. For $K_{ATP}$ channel, although the detailed role of free energy is not clear, research are undergoing on this topic\cite{nicholls} and ATP hydrolysis is demonstrated to be relevant. We believe that free energy is important for its proper function.
	\par
	Experiments remain to be done to demonstrate our main idea that $\Delta G$ dominates the bifurcation process. To achieving this, [ATP] and $\gamma$ should be monitored simultaneously as well as calcium signal. The environments in each cell vary, so the bifurcation border should be get by statistical analysis of multiple data. Then it's straight forward to show if the border is sensitive to $\Delta G$.
	\par
	Some experiment indicates that there is connection between aging and diabetes.\cite{li_defects_2014} $^,$\cite{barker2015beta} Although we believe that $\Delta G$ of ATP hydrolysis remains approximately constant in normal physiological environment, in aged cell or abnormal cell, the ability to generate ATP and sustain a high $\Delta G$ level decreases. This leads to a difficulty in calcium oscillation. It is also possible that a parameter's change causes shift of the bifurcation border, resulting in similar difficulty. Our research indicates a underlying mechanism of type II diabetes which has a higher incidence among the aged people.
	\par
	There are new experimental data showing that ATP would drop with oscillation of calcium.\cite{li2013oscillations} There would be no oscillation where ATP is highest. However, our model can not simulate the baseline of calcium, which is low with highest ATP. Since our model derives from the Bertram model, this indicates that our model is still not complete. Nevertheless, $\Delta G$ may still provide some insight over it. In a given adenosine pool, ATP:ADP ratio would change significantly when ATP is about filling the pool. The increase of $\gamma$ at highest ATP may stimulate ATPases, yielding the lower calcium baseline as in the "dip" area.\cite{barker2015beta}

\section*{Acknowledgments}
We thanks for Dr. Liangyi Chen, Dr. Chao Tang, Huixia Ren, and Chensheng Han for helpful discussions and comments.

	\begin{table}[H]
		\centering
		\caption{Parameter setting in our model. Parameters marked as Ad are adjusted parameters, whose sensitivity analysis are discussed in Appendix-II. The values of most parameters are adopted from the Bertram's model, which are similar to those in other models in essence.}
		\begin{ruledtabular}
		\begin{tabular}{ccccc}
			Model Parameters & Value & Units & Eq.   & Reference \\
			$Pi$  & 1000  & $\mu M$ & \ref{deltag},\ref{gamma} & \cite{qian} \\
			$K_{eq}$ & $4.9 \times 10^{11}$ & $\mu M$ & \ref{gamma} & \cite{qian} \\
			$k_{pmca}$ & 4     & $\mu M$ & \ref{JPMCA} & Ad \\
			$G_{K(ATP)}$ & 25000 & pS    & \ref{IKATP} & \cite{Bertram} \\
			$g_{K(Ca)}$ & 600   & pS    & \ref{IKCa} & \cite{Bertram} \\
			$V_K$ & -75   & mV    & \ref{IKCa},\ref{IKATP},\ref{IK} & \cite{Bertram} \\
			$V_{Ca}$ & 25    & mV    & \ref{ICa} & \cite{Bertram} \\
			$g_K$ & 2700  & pS    & \ref{IK} & \cite{Bertram} \\
			$g_{Ca}$ & 1000  & pS    & \ref{ICa} & \cite{Bertram} \\
			$C_m$ & 5300  & fF    & \ref{potential} & \cite{Bertram} \\
			$k_D$ & 0.5   & $\mu M$ & \ref{omega} & \cite{Bertram} \\
			$k_{dd}$ & 17    & $\mu M$ & \ref{theta} & \cite{magnus} \\
			$k_{td}$ & 26    & $\mu M$ & \ref{theta} & \cite{magnus} \\
			$k_{tt}$ & 1     & $\mu M$ & \ref{theta} & \cite{magnus} \\
			$v_m$ & -20   & mV    & \ref{minfty} & \cite{Bertram} \\
			$v_n$ & -16   & mV    & \ref{minfty} & \cite{Bertram} \\
			$s_m$ & 12    & mV    & \ref{minfty} & \cite{Bertram} \\
			$s_n$ & 5     & mV    & \ref{minfty} & \cite{Bertram} \\
			$[Ca^{2+}]_{pla}$ & 1000  & $\mu M$ & \ref{JPMCA} & \cite{dupont} \\
			$k_1$ & 1     & $\mu M^{-3} \ ms^{-1}$ & \ref{JPMCA} & Ad \\
			$k_{-1}$ & $8 \times 10^{-8}$ & $\mu M^{-1} \ ms^{-1}$ & \ref{JPMCA} & Ad \\
			$k_2$ & 5     & $\mu M \ ms^{-1}$ & \ref{JPMCA} & Ad \\
			$\alpha$ & $4.5 \times 10^{-6}$ & $fA_{-1}\mu M\ ms{-1}$ & \ref{Cacyto} & \cite{Bertram} \\
			$f_{cyt}$ & 0.01  &       & \ref{Cacyto} & \cite{Bertram} \\
		\end{tabular}%
		\end{ruledtabular}
		\label{tab:pt}%
	\end{table}
	
\end{document}